\documentclass[12pt,aps,prl,reprint,groupedaddress,onecolumn]{revtex4-1}

\usepackage{graphicx}
\usepackage{color}

\begin{document}



\title{Twist neutrality, a zero sum rule for oriented closed space curves\\ 
with applications to circular DNA}


\author{Jakob Bohr}
\email[]{jabo@nanotech.dtu.dk}
\affiliation{DTU Nanotech, Building 345\O, \O rsteds Plads, Technical University of Denmark, 2800 Kongens Lyngby, Denmark\\}

\author{Kasper W. Olsen}
\email[]{kasol@nanotech.dtu.dk}
\affiliation{DTU Nanotech, Building 345\O, \O rsteds Plads, Technical University of Denmark, 2800 Kongens Lyngby, Denmark\\}


\date{\today}

\begin{abstract}
\vspace{0.8cm}
The interplay between global constraints  and local material properties of chain molecules is a subject of emerging interest. Studies of molecules that are intrinsically chiral, such as double-stranded DNA, is one example. Their properties generally depend on the local geometry, i.e. on curvature and torsion, yet the paths of closed molecules are globally restricted by topology.  Molecules that fulfill a twist neutrality condition, a zero sum rule for the incremental change in the rate of winding along the curve, will behave neutrally to strain. This has implications for plasmids. For small circular microDNAs it follows that there must exist a minimum length for these to be double-stranded. It also follows that all microDNAs longer than the minimum length must be concave. This counterintuitive result is consistent with the kink-like appearance which has been observed for circular DNA. A prediction for the total negative curvature of a circular microDNA is given as a function of its length.
\end{abstract}


\maketitle

\section{}

Here, a simple topological sum rule for closed space curves will be applied to a physical model where the molecular properties arise by local interactions. The motivation is to better understand what determines the shape of closed loops of chain molecules, e.g. DNA. Intuitively, the double-stranded helical geometry of DNA reveals that topological restrictions apply and suggests that one can expect the existence of fascinating phenomena. One example hereof was recently reported in a study of the supercoiled DNA that forms plectonemes, where long-ranged dynamical effects involving two distant plectonemes were observed \cite{loenhout2012}. Theoretical aspects of the physics of polymers and DNA have been reviewed in detail \cite{frank1981,micheletti2011}. An early triumph is the White's theorem, also associated with C\v{a}lug\v{a}reanu and Fuller \cite{white1969,calugareanu1961,fuller1971,pohl1968}, that imposes powerful restrictions for the ways a closed ribbon can change geometry, and ways it cannot change geometry. The theorem describes the relationship between linking, twist and writhe, i.e. $Lk=Tw+Wr$. The White's theorem does not say anything about the material properties of DNA. For example, it does not say whether the two strands prefer to be straight, or in a helical configuration, relative to each other. The DNA has a molecular preference for the right-handed double helix (B-DNA). This preference breaks the symmetry and makes DNA chiral, which gives it special properties. For example, a coupling between strain and twist that we will assume to be local as it depends on short-ranged molecular forces. Such properties therefore depend on the local geometry. In most cases global topological requirements and local geometrical preferences will be in conflict. In the following we will focus on the case where they are concurrent.\\ 

For two three-dimensional space curves, $\alpha$ and $\beta$, that are closed and oriented their Gauss linking number, $Lk(\alpha,\beta)$, is a topological conserved integer \cite{ricca2011}. The linking number describes the number of times the two curves interlink each other. The self-linking number, $Lk(\alpha, \vec{N})$, of a regular space curve, $\alpha$, requires a differentiable field of unit normal vectors, $\vec{N}$. In the following we assume the curve to be smooth to the order of at least $\mathcal{C}^3$. Together, the unit tangent, the unit normal, and the unit binormal vectors frames the curve. The White's theorem applies equally well to framed curves and the total twist becomes the twist integral of the vector field $\vec{N}$ besides a factor of $2\pi$. Another number which describes the vector field $\vec{N}$ is the winding number. The winding number requires to be measured relative to something, i.e., we must have two vector fields defined at the same time. Here we will use the Frenet frame $\vec{N}_F$, actually a modified Frenet frame, $\vec{N}_{mF}$, 
\cite{mod1}, as reference.  
It later becomes a physically relevant frame for winding comparisons of intrinsically chiral molecules with non-chiral molecules.\\ 

Let $\theta$ be the angle between the two local frames, i.e. $\theta = \arccos(\vec{N} \cdot \vec{N}_{mF})$, then the winding number, $\Phi$, of $\vec{N}$ relative to the modified Frenet frame becomes

\begin{equation}
\label{eq:winding}
\Phi =\frac{1}{2\pi} \int_\alpha \frac{\partial \theta}{\partial s} {\rm d}s \, .
\end{equation}

\noindent One could have chosen another reference frame, for example the "Dennis-Hannay-Maddocks" frame \cite{dennis2005}. With this type of reference frame the winding number of the normal vector field becomes equal to its linking number. With the modified Frenet frame as reference, the winding number becomes

\begin{equation}
\label{eq:winding2}
\Phi = Lk(\alpha,\vec{N})-Lk(\alpha,\vec{N}_{mF}) \, .
\end{equation}

\noindent Therefore $\Phi$ is a topological conserved integer. Let us now consider a bundle of frames, $\vec{N}(u)$, $u \in \bf{R}$, which are differentiable in the vicinity of $u=0$ and for which $\vec{N}(0)=\vec{N}$. It follows from differentiation of Eq. (\ref{eq:winding}) that 

\begin{equation}
\label{eq:lemma}
\int_\alpha \frac{\partial^2 \theta(s,u)}{\partial u \partial s}  {\rm d}s= 0 \, .
\end{equation}

\noindent This is a remarkable simple result that we denote the twist neutrality lemma. In words, the lemma states that the differential changes in the rate of the winding progression must obey  a zero sum rule.\\

Double-stranded DNA is a helical molecule and it therefore has a chiral symmetry, hence some of the properties of DNA will reflect this symmetry, e.g. its response to strain. For comparisons of the properties of chiral DNA to a hypothetical DNA with no intrinsic chirality it  will be useful to choose for the space curve the central line of the double-stranded DNA. Then the framing can be described by the helical path of one of the two strands. By placing a ribbon containing the two strands and the central lines of the helical DNA one can see that the linking of the central line with one of the strands is equal to the linking of one strand with the other \cite{mod2}.\\ 

In the following we will make the assumption that the strain-twist coupling for a short stretch of DNA only depends on its local geometry, i.e.~on its strain, curvature and torsion $(\sigma,\kappa,\tau)$. Hence, at small strains the incremental change in the rate of winding, $\delta$, will depend only on $\kappa$ and $\tau$,

\begin{equation}
\label{eq:ratewinding}
\delta (\kappa,\tau) = \frac{\partial^2 \theta}{\partial \sigma \partial s} |_{\sigma =0} \, .
\end{equation}

\noindent The physics which underlines this assumption is the relatively local nature of the  interactions within DNA between its constituent atoms. After all it is their interactions that determine the molecular structure as well as its reaction to strain. In Eq. (\ref{eq:ratewinding}) we have chosen not to make the orientation of the double helix, i.e. $\theta$, enter the equation. This means that material properties are considered uniform over about a pitch of the double helix which is $\sim$10 base pairs (bp) long. From applying the twist neutrality lemma to strain we now have 

\begin{equation}
\label{eq:lemma2}
\int_{\rm DNA} \delta(\kappa,\tau)  {\rm ~d}s= 0 \, .
\end{equation}

\noindent This restriction, to be twist-neutral in the sense described, is a severe constraint that significantly reduces the number of possible closed space curves that double-stranded DNA can make. They are not only topological restricted by the White's theorem, they are also geometrically restricted because of Eqs. (\ref{eq:ratewinding}) and (\ref{eq:lemma2}). Thus, strained non-conforming paths would force the DNA to depart from Eq. (\ref{eq:lemma2}). In the worst case, it could break one or both strands of the DNA molecule, in milder cases it could lead to a local detachments of the two strands from each other, i.e. local melting of the DNA.
Denaturation bubbles have been observed in atomic force microscopy (AFM) images of circular DNA \cite{jeon2010,adamcik2012} and their closure dynamics has been studied by dynamics simulations \cite{dasanna2012}.\\

The implications of twist neutrality can easily be addressed for closed planar curves such as plasmids deposited on a substrate. Long plasmids can display many crossings such as supercoils and plectonemic structures \cite{lee2012}. For shorter plasmids cooperative kinking at distant sites has been considered \cite{lionberger2011}. 
The simplest cases are planar plasmids with no self-crossings, i.e. the centerline of the DNA is topologically equivalent to a circle. Recall that straight DNA winds when stretched. It rotates opposite of unwinding, a behavior that sometimes is denoted overwinding \cite{lionnet2006,gore2006}. In the language of this paper it means that
$\delta > 0$. A straight section of DNA can therefore not satisfy the twist neutrality criterion. Well, that is a hypothetical statement as a straight line it is not a closed curve, neither! Can a circle of DNA fulfill the twist neutrality criterion? As DNA bends, we will assume that $\delta$  decreases monotonically as a function of the curvature describing this bending and that $\delta$ eventually becomes negative. 
This behavior is also observed for tubular double helices \cite{olsen2012}, while the opposite behavior, i.e. for $\delta$ to increase monotonically, would not be possible as we know that DNA can form closed loops. For a perfect geometrical circle to fulfill the twist neutrality criterion it must 
have $\delta=0$ everywhere, which is only fulfilled for a specific radius $R_0$. It means that plasmids that are truly circular, and not just called circular because they are loops, come in one specific size with the plasmid length $L_0=2\pi R_0$. Consider a plasmid of length $L$  different from $L_0$. We series expand $\delta$ to include the linear turn, i.e.,

\begin{equation}
\delta= \lambda ( | \kappa |-\kappa_0) \, ,
\end{equation}

\noindent where $\kappa$ is the curvature with sign as conventionally used for two-dimensional curves, $\kappa_0=1/R_0$ is the curvature of the circle of radius $R_{0}$, and $\lambda$ is a constant that determines the coupling between twist and curvature. 
{\color{black} Since the curve is assumed planar, $\delta$ is independent of the torsion, $\tau$.
}
The twist neutrality criterion then becomes

\begin{equation}
\label{eq:lemma3}
\int_{0}^L (| \kappa(t)| -\kappa_0) \frac{{\rm d}s}{{\rm d} t} {\rm d}t = 0 \, ,
\end{equation}

\noindent where $\kappa(t)$ is the curvature measured along the plasmid, and $s(t)$ the corresponding arc length. Therefore, we obtain

\begin{equation}
\label{eq:totalcurv}
\int_0^{L} | \kappa | ds = 2\pi\frac{L}{L_0} \, .
\end{equation}

\begin{figure}[t]\centering
\includegraphics[width=14.2cm]{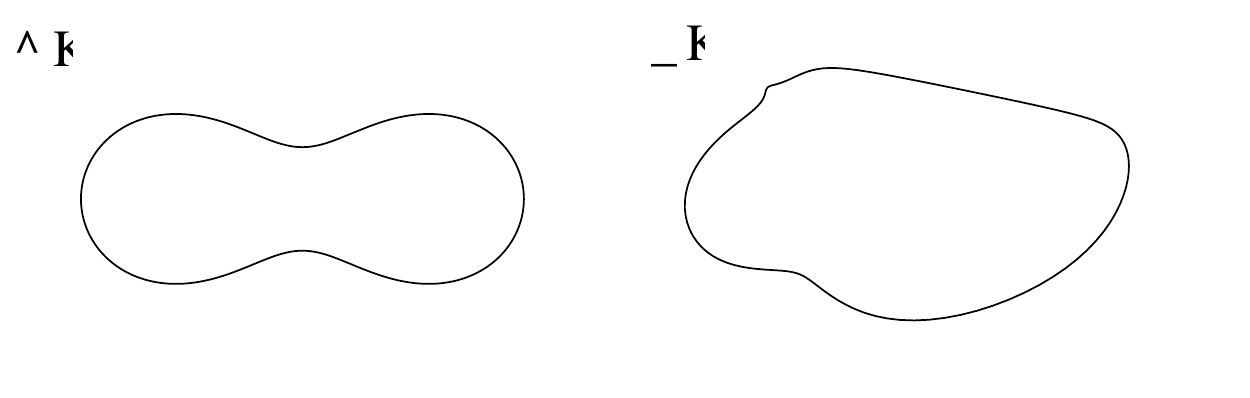}
\caption{\it Examples of non-convex solutions to the twist neutrality condition for (A) $L/L_0=1.50$, and (B) $L/L_0=1.46$. There are no solutions with $L/L_0 <1$.}
\end{figure}

\noindent In two dimensions the integral in Eq.~(\ref{eq:totalcurv}) over a closed curve is always greater than or equal to $2\pi$ \cite{milnor1950}. Consequently, there are no possible solutions for plasmids with a length shorter than $L_0$. Shorter plasmids seem to be required to be linear, or to melt and become circular single strands, which presumably themselves have yet a smaller lower cutoff for their circumference.\\

What are the possible shapes of plasmids which satisfy the "first-order" twist neutrality criterion Eq. (\ref{eq:totalcurv}) with a length $L$ larger than the smallest plasmid length, $L_0$? The shape must be concave as the integral of the total curvature equals $2\pi$ exactly when the plasmid is convex \cite{milnor1950}. 
To find a possible shape we cannot use a circle or an ellipse as they are both convex. One concave curve with a simple mathematical description is the {\it Cassini oval}, which is given by the polar expression:

\begin{equation}
r^4 - 2a^2r^2 \cos (2\theta) = b^4 - a^4\, ,
\end{equation}

\noindent where $a$ is half the distance between the two foci, and $b$ the square root of the product of the distances to the two foci. 
The shape of the Cassini ovals depends on the ratio $\epsilon=a/b$. For $\epsilon<1$ the shape is a single loop with an oval (convex) or peanut (concave) shape, for $\epsilon=1$ the lemniscate is produced.
Figure 1A shows a peanut-shaped Cassini curve which is a solution to Eq. (\ref{eq:totalcurv}) with $L=1.5 L_0$. The parts of the Cassini oval where the curvature is negative (the progression of the curve is clockwise) are clearly visible. However, the Cassini ovals are not good descriptors of circular DNA which generally are more random in appearance. The desire for some amount of randomness was demonstrated in an AFM study of the bond correlation function \cite{witz2008}. In this AFM study it was shown that the halfway bond correlation was not completely antiparallel, a result which infers some amount of randomness. By a simple random procedure choosing 15 points and generating a closed curve with a spline, a somewhat more realistic (though not perfect) concave curve is drawn, see Fig. 1B. For this curve $L=1.46 L_0$. The concave nature of closed plasmids is also consistent with the existence of the type of kinks observed in circular DNA \cite{han1997,zhao1998}.\\

For the general case of a concave curve which satisfies Eq. (\ref{eq:totalcurv}) one has,

\begin{equation}
\label{eq:mcurv}
I_{-}=\int_0^L \kappa_{-}{\rm ~ }ds = \frac{L-L_0}{L_0} \pi\, .
\end{equation}

\noindent The integral is the total negative curvature without sign, where $\kappa_{-}$ is equal to $| \kappa |$ when $\kappa<0$, otherwise $\kappa_{-}$ is zero. 
This shows that the total negative curvature must be a linear function of the length, $L$, of circular DNA. A prediction that can be experimentally tested by plotting $I_{-}$ versus $L$, e.g. in a scatter plot.
Likewise, using a corresponding definition for $\kappa_+$, one can calculate a total positive curvature, 

\begin{equation}
I_{+}=\int_0^L \kappa_{+}{\rm ~ }ds = 2\pi +\frac{L-L_0}{L_0} \pi\, ,
\end{equation}

\noindent hence $I_{+}-I_{-} =2\pi$ and $I_{+}+I_{-} =2\pi L/L_0$.
One way to estimate the value of $L_0$ would be to invoke molecular modeling of DNA on a sufficiently large scale to study the influence of curvature. Tubular models offer a way of getting a first attempt of an estimate. Recently, we found that a tubular model of B-DNA has no strain-twist coupling for a circular ring structure with a diameter of  $2R= 156$~\AA ~\cite{bohr2012}. The corresponding estimate of $L_0$ is about 124 bp. Within the tubular model the uncertainty of this result is at most 5-10 \%, however, a tubular model may not fully capture all numerical aspects of the molecular structure of DNA and the real discrepancy may well be larger. The above estimate for the lower bound, $L_0$, is in reasonable agreement with lengths recently found for small loops of double-stranded DNA in a study of extrachromosomal microDNAs \cite{shibata2012}. They identified tens of thousands of short circular DNA in mouse tissue as well  mouse and human cell lines with the overall size distribution in the range 80 to 2000 bp with a strong propensity for lengths peaked at around 150 and 300 bp for the human DNA and at 180, 360, 540, 720 and 900 bp (plus perhaps two more peaks) for mouse DNA. Circular single-stranded loops were reported as being yet shorter. The apparent tendency for an integer multiple of a preferred length can in the perspective of the above analysis be suggestive of microDNAs that coil multiple times when {\it in vivo} or are forming plectonemic conformations. Alternatively, it could also have other origins such as the specific base pair sequence.\\

Theoretically, the regime of relatively stiff polymers for which the elastic response prevails has been considered \cite{alim2007,ostermeir2010}. At biological temperatures DNA is fairly close to the melting temperature of the double strand; the melting temperature depends in detail on such properties as the salt concentration of the aqueous solution and on the sequence of the DNA. Therefore, thermal fluctuations in DNA  are omnipresent. Fluctuations in DNA mini-plasmids have been modeled as a Kirchhoff-Clebsch rod with no strain \cite{tobias1998}. Twist neutrality engender differentiable frames for circular DNA, which is suggestive of the existence of well-behaved excitations and fluctuations, e.g. breathing-modes. It would be interesting to include strain and stretching in detailed studies of the different kinds of modes and their stability \cite{dempsey1996,zakrzhevskii2011}, and to include twist neutrality in twist-stretch elasticity studies of closed loops of chiral filaments \cite{upmanyu2008}. It would also be interesting to include higher-order curvature terms in the twist neutrality condition. This includes second-order terms which are also found in bending energy and elastica \cite{langer1984}.
\\

In summary, a twist neutrality perspective has been given on molecules which display a strain-twist coupling reflecting the chiral nature of the molecule. For circular plasmids is was shown that all such plasmids except the one of minimal length, $L_0$, must have a concave shape. 
The predictions of Eq. (\ref{eq:mcurv}) can be experimentally tested by measurements of the total negative curvature. Further, this will allow for a self-consistent estimate of $L_0$.\\

JB thanks the Isaac Newton Institute for hosting the inspiring workshop "Topological Aspects of DNA Function and Protein Folding". We would like to thank Giovanni Dietler and Ralf Metzler for references. This work is supported by the Villum Foundation.


\begin{thebibliography}{99}
\bibitem{mod1} The modified Frenet frame $\vec{N}_{mF}$: The Frenet frame requires curvature to be everywhere non-zero. This unnecessarily restricts the class of curves which can be considered. Instead we will consider a modified Frenet frame where the curvature is allowed to be zero in discrete points. The framing is then differentially patch together from sections of Frenet frames, and sections where the normal vector points in the opposite direction of $\vec{N}_F$, whichever provides a differential framing. 
\bibitem{mod2} This is under the assumption that the double-stranded DNA does not intercept itself.
\end{thebibliography}
\end{document}